\documentclass[a4paper,fleqn,usenatbib]{mnras}

\usepackage{newtxtext,newtxmath}

\usepackage[T1]{fontenc}
\usepackage{ae,aecompl}

\usepackage{graphicx}
\usepackage{amsmath}	
\usepackage{amssymb}	
\usepackage{xcolor}

\title[Time variable radio emission from a T-Tauri flare]{Predicting the time variation of radio emission from MHD simulations of a flaring T-Tauri star}

\author[C. O. G. Waterfall et al.]{
C. O. G. Waterfall,$^{1}$\thanks{E-mail: charlotte.waterfall@postgrad.manchester.ac.uk} 
P. K. Browning,$^{1}$
G. A. Fuller,$^{1}$
M. Gordovskyy,$^{1}$
\newauthor
S. Orlando,$^{2}$
F. Reale$^{3,2}$
\\
$^{1}$Jodrell Bank Centre for Astrophysics, Department of Physics and Astronomy, School of Natural Sciences, The University of Manchester, Manchester, M13 9PL, UK\\
$^{2}$ INAF - Osservatorio Astronomico di Palermo, Piazza del Parlamento 1, I-90134, Palermo, Italy\\
$^{3}$ Dipartimento di Fisica \& Chimica, Università di Palermo, Piazza del Parlamento 1, I-90134 Palermo, Italy
}

\date{Accepted XXX. Received YYY; in original form ZZZ}

\pubyear{2019}

\begin{document}
\label{firstpage}
\pagerange{\pageref{firstpage}--\pageref{lastpage}}
\maketitle

\begin{abstract}
\\ 
We model the time dependent radio emission from a disk accretion event in a T-Tauri star using  3D, ideal magnetohydrodynamic simulations combined with a gyrosynchrotron emission and radiative transfer model. We predict for the first time, the multi-frequency (1-1000\,GHz) intensity and circular polarisation from a flaring T-Tauri star. A flux tube, connecting the star with its circumstellar disk, is populated with a distribution of non-thermal electrons which is allowed to decay exponentially after a heating event in the disk and the system is allowed to evolve. The energy distribution of the electrons, as well as the non-thermal power law index and loss rate, are varied to see their effect on the overall flux.  Spectra are generated from different lines of sight, giving different views of the flux tube and disk. The peak flux typically occurs around 20$-$30\,GHz and the radio luminosity is consistent with that observed from T-Tauri stars. For all simulations, the peak flux is found to decrease and move to lower frequencies with elapsing time.   The frequency-dependent circular polarisation can reach 10$-$30$\%$ but has a complex structure which evolves as the flare evolves. Our models show that observations of the evolution of the spectrum and its polarisation can provide important constraints on physical properties of the flaring environment and associated accretion event. \end{abstract}



\section{Introduction}
T-Tauri stars are dynamic and variable low mass objects, with classical T-Tauri stars in particular hosting large and dense circumstellar disks \citep{feigelson1999high}.  They frequently exhibit flaring-like activity that generates significant levels of radio and X-ray emission.  This emission is thought to be a result of the reconnection of magnetic field lines, either within the inner magnetosphere of the star itself or from interaction of the stellar and accretion disk's field lines \citep{Osten2009}. This is similar to the behaviour seen in emission signatures from solar flares (\citet{Benz2008}, \citet{Fletcher2011} and more recently by \citet{fleishman2020decay}) although on a considerably larger scale. Observations of the radio (and X-ray) emission from these pre-main sequence stars can probe the physical environment and processes involved in the star-disk interaction, which includes its role in transferring material from the disk to the star,  driving stellar outflows and extracting the angular momentum from the system. By modelling flaring emission we can learn more about these environments, as well as enabling the comparison of results to flares on other stars in addition to T-Tauri observations.

\begin{figure}
	\includegraphics[width=\columnwidth]{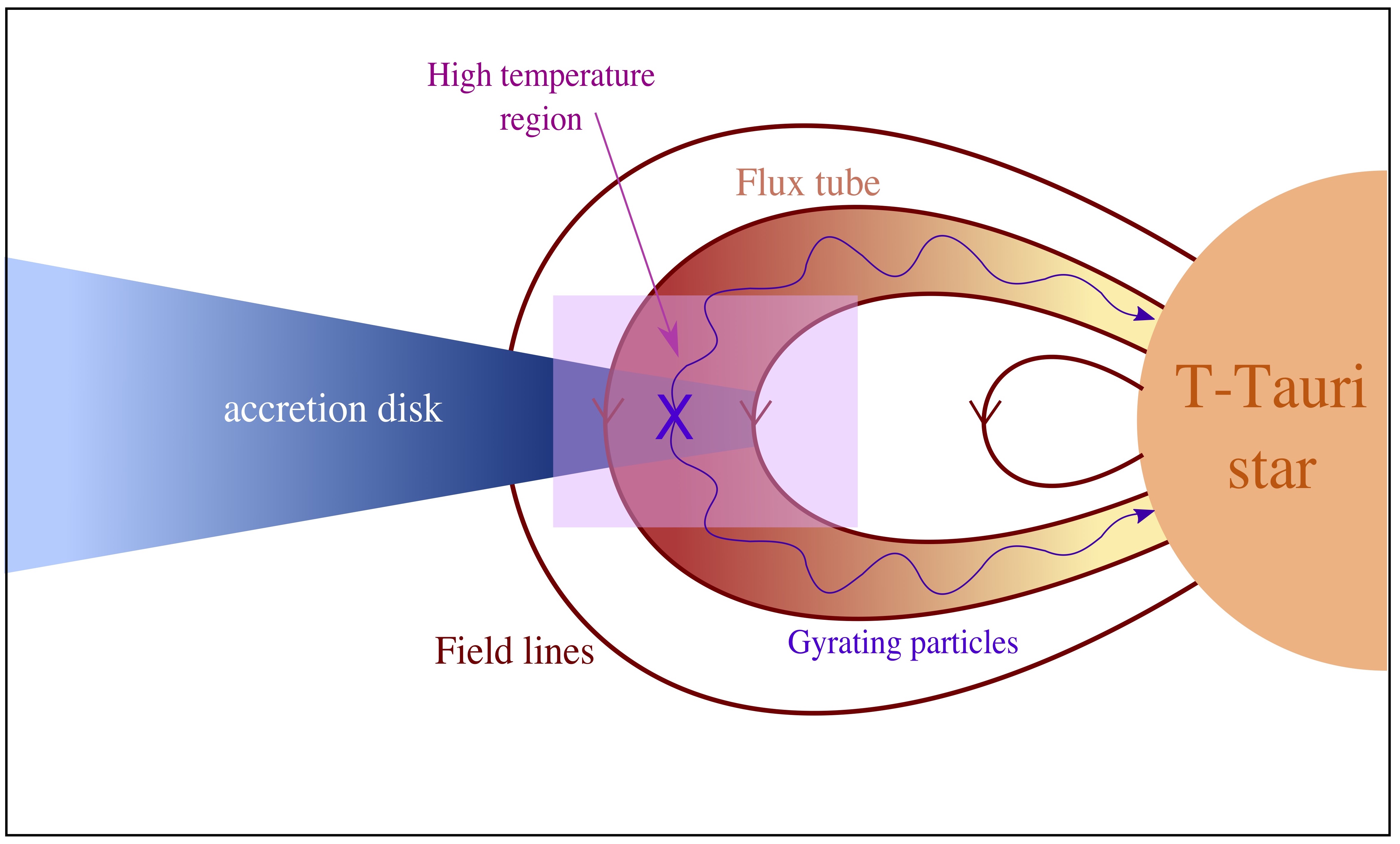}
    \caption{Schematic of a flaring flux tube connecting a star and its circumstellar disk.  Magnetic reconnection occurs at the labelled X-point between the field lines of the disk and star.  Non-thermal particles gyrate from this point towards the stellar surface along the field lines.  The highest particle density is at this loop top in the high temperature region.}
    \label{fig:star}
\end{figure}

The non-thermal radio and thermal X-ray emission generated in T-Tauri flares was explored by \citet{waterfall2018modelling} (hereafter referred to as Paper I).  In that work a flux tube connecting a T-Tauri star to its disk was defined and populated with non-thermal and thermal particles, within a background hydrostatic coronal atmosphere. The 3D GX simulator \citep{Nita2015}, which calculates gyrosynchrotron emission and absorption using fast algorithms from \citet{fleishman2010fast}, was used to calculate the X-ray and radio emission generated from this `flaring' flux tube.  A multipolar (dipole plus octupole) magnetic field was used and the parameters were varied to give a range of possible X-ray and radio luminosities that might be observed. The typical luminosities from this model were logL${_\mathrm{X}}$ (erg s$^{-1}$) = 30.5, logL${_\mathrm{R}}$ (erg s$^{-1}$ Hz $^{-1}$) = 16.3. These luminosities are several orders of magnitude higher than those obtained from solar flares, as expected from these active T-Tauri stars.  The flux tube used was over $10^{11}$cm in length, containing plasma temperatures and densities that were also larger than typical solar flare values \citep{shibata1999origin,shibata2011solar}. However, the modelling did not explore the effect of time on the emission. In contrast with Paper I, we here use a time dependent 3D magnetohydrodynamic (MHD) model of a flaring atmosphere and accretion disk system \citep{orlando2011mass}.  The model follows accretion onto a star, induced by a flaring event that triggers the development and evolution of a flux tube that is populated with non-thermal electrons. An overview of this set up is shown in Fig.~\ref{fig:star}.

Accretion and flaring in young stars have previously been investigated through MHD modelling: \citet{orlando2013radiative} modelled accretion shocks onto the surface of classical T-Tauri stars that are generated from plasma moving along non-uniform field lines.  They performed 2D ideal MHD simulations and found that the magnetic field properties heavily influenced some of the plasma's properties.  

Other models investigate accretion onto young stars using time-dependent MHD with different magnetic field configurations, such as \citet{long2012accretion} and \citet{robinson2017time}.  \citet{long2012accretion} used global 3D MHD simulations to model stellar accretion for stars with multipolar magnetic fields.  \citet{robinson2017time} also looked at multipolar field configurations, as well as purely dipole fields for comparison, through a 1D time-dependent model. 

Some previous work has also considered the emission generated from these events, such as \citet{sacco2008x}.  They performed 1D hydrodynamical (HD) simulations to model the effect of accretion onto the chromosphere and the resultant X-ray emission and densities that are produced.  \citet{sacco2010observability} again modelled this accretion through flux tubes connecting the star and disk and discussed the X-ray emission and observations from these regions. The idea of a flaring loop connecting an inner disk to its young star via a magnetic flux tube was also explored in 1D hydrodynamic modelling performed by \citet{isobe2003hydrodynamic}.

X-ray and radio emission are key signatures of solar and stellar flares occurring and are used in tandem to define the G\"udel-Benz relation \citep{Guedel1993}.  This relationship correlates these two emissions generated in the coronae of magnetically active stars:

\begin{equation}
    \frac{L_{\rm{X}}}{L_{\mathrm{R}}} \approx 10^{15.5}
\label{eq:relation}
\end{equation}

where L${_\mathrm{X}}$ and L${_\mathrm{R}}$ are the X-ray and radio luminosity respectively.  The relation holds for a range of flares, from solar microflares up to large main sequence flares.
 
This relationship was extensively explored for T-Tauri stars in Paper I.  Both the X-ray and radio luminosity of observed flares as well as a range of modelled flares were compared to the relationship and found to lie under it, indicating an overproduction of radio emission. This provides a useful tool for assessing the predictions of models, and we will revisit this with the results of the present model.  However, our main goal here is to develop time-dependent predictions of observing radio signatures from gyrosynchrotron emission due to non-thermal electrons and heated plasma, based on a realistic 3D model of a flaring atmosphere.  We will consider how a flare might appear from different lines of sight, predicting intensities in different frequencies as well as degrees of circular polarisation. 

Results from a 3D MHD model \citep{orlando2011mass} are used along with a flux tube containing non-thermal particles in calculating the gyrosynchrotron emission from a flaring T-Tauri star.  The background of this MHD model, discussion of the flux tube (and non-thermal particles) and a description of the fast GS simulations used to calculate the emission is given in Section~\ref{sec:models}. The results of these simulations will be discussed in Section~\ref{sec:results} and the effect of varying some of the model components such as the power law index and distribution of particles is explored in Section~\ref{subsec:vary}. The results are summarised and discussed in Section~\ref{sec:conc}.

\section{The models}
\label{sec:models}
In Paper I we used the GX simulator to calculate the thermal and non-thermal radio and X-ray emission from a flaring loop approximating that from a T-Tauri star (with mass and radius equal to that of the Sun) \citep{Nita2015}. The GX simulator has been designed to enable three-dimensional modelling of the emission from solar flares and is downloadable via the solar software (SSW) distribution website: \url{http://www.lmsal.com/solarsoft/ssw_install.html}. The GX simulator is based on fast gyrosynchrotron codes developed by \citet{fleishman2010fast}.  These codes work by calculating the gyrosynchrotron emission along one-dimensional lines of sight. It is these original codes that are adapted here to model the radio emission from the numerical magnetohydrodynamic model of a heating event in a circumstellar disk around a T-Tauri star. These codes (hereafter referred to as the fast GS simulations) have been adapted to cover an entire simulation, integrating across multiple lines of sight. The fast GS simulations allow us more flexibility over the 3D GX simulator, for example allowing the use of the time evolving 3D physical properties (density, temperature and magnetic field structure) from the MHD simulation of \citet{orlando2011mass} to model the emission from a T-Tauri star over several hours. These fast GS simulations allow us to produce spectra and calculate gyrosynchrotron emission from a predefined flaring flux tube, using a semi-empirical model of the non-thermal electrons within the tube (Section~\ref{subsec:fieldline}).

\subsection{3D MHD model}
\label{subsec:mhd}
The data from \citet{orlando2011mass} describes the modelling of a
flaring event in the magnetosphere of a rotating classical T-Tauri
star that leads to accretion from the disk onto the stellar surface.
The 3D MHD model involves a central star with mass and radius of
$M=0.8\,M_{\odot}$ and $R=2\,R_{\odot}$, respectively, surrounded by a thick
quasi-Keplerian disk. The star rotates with a period of 9.2\,days about
an axis coincident with the normal to the disk mid-plane and has a
dipole like field aligned with the rotation axis and of strength $\approx
1$~kG at the stellar surface. Initially the disk is dense, cold,
and isothermal with a temperature of $8\times 10^3$~K. The disk is
truncated by the stellar magnetosphere at $R_{\rm t}
=2.86\,R_{*}$. The corotation radius, where a Keplerian orbit
around the star has the same angular velocity as the star’s surface,
is $R_{\rm co} = 9.2\,R_*$. The corona of the star initially is
isothermal with a temperature of 4~MK and density $10^8-10^9$~cm$^{-3}$.

\begin{figure}
	\includegraphics[width=0.95\columnwidth]{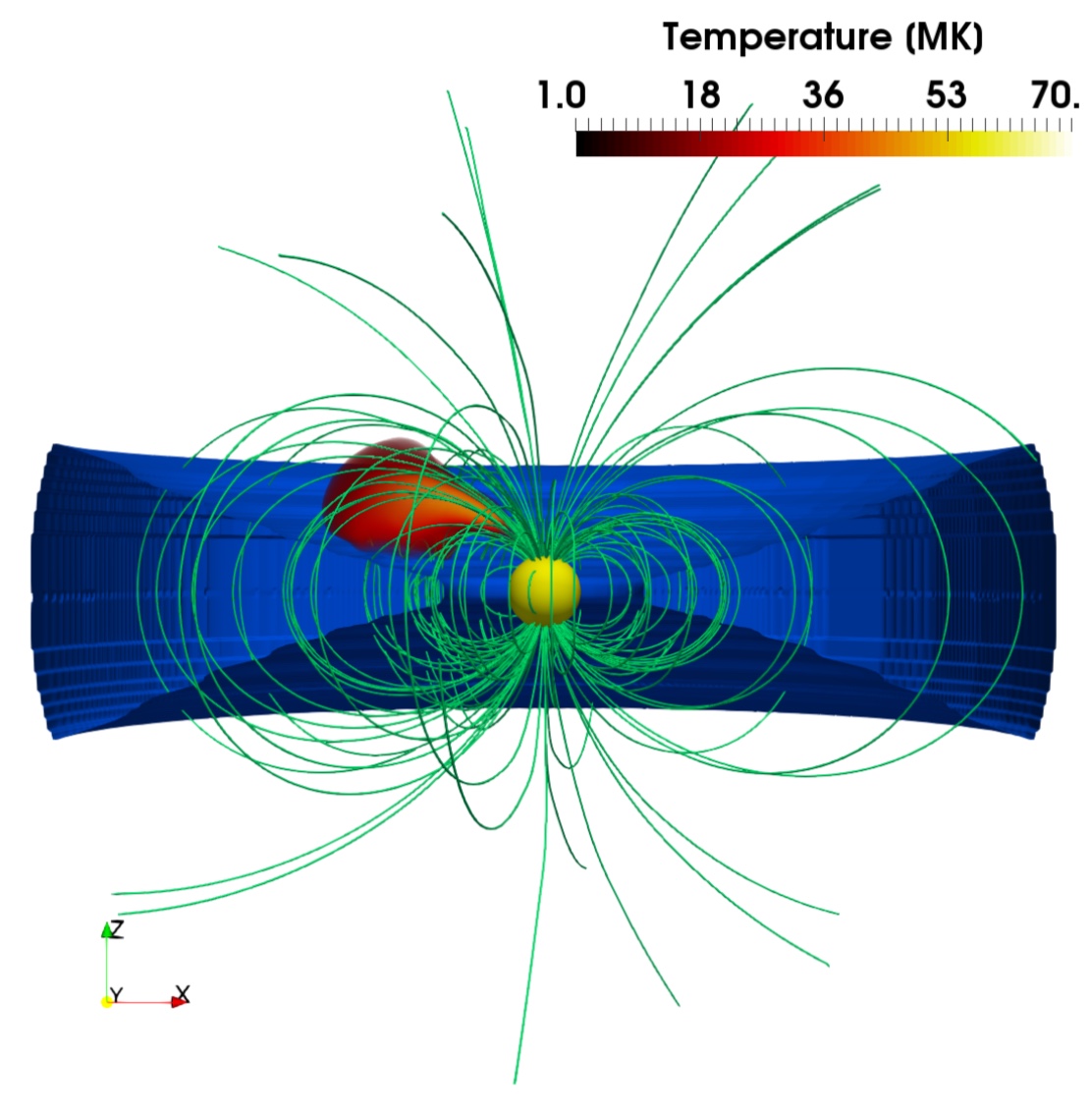}
    \caption{Edge on view of the central star surrounded by its accretion disk at t = 1.2 hours.  The hot flaring loop that has developed joins these two regions. The dipolar magnetic field configuration at this time is shown in green. }
    \label{fig:domain}
\end{figure}

The model takes into account the gravitational force, the disk viscosity,
the magnetic-field-oriented thermal conduction (including
the effects of heat flux saturation), the radiative losses from
optically thin plasma, and the coronal heating (including a component describing the stationary coronal heating, plus a transient component triggering a flare). The flaring event is created by initiating a
transient heat pulse that is switched off after 300 seconds and allowed to
evolve. This heat pulse is released at a distance of $5\,R_*$ from the stellar surface, just outside the surface of the accretion disk, namely well below the corotation radius
$R_{\rm co}$. The heat propagates along the field lines connecting the disk to the star, thus creating a hot flaring loop between the star and heated region. The star-disk system and the
flaring loop at $t = 1.2$~hours are illustrated in Fig.~\ref{fig:domain}. The three
lines of sight that will be considered, along the X, Y and Z axis, are also shown. The Z line of sight looks down onto the disk, in the XY plane whereas the X and Y lines of sight look side on through the disk. We take a general approach using these three lines of sight, rather than selecting a specific inclination for a particular T-Tauri star. Modelling of any observed T-Tauri star in the future should account for the specific inclination with respect to the line of sight.

\begin{figure}
	\includegraphics[width=\columnwidth]{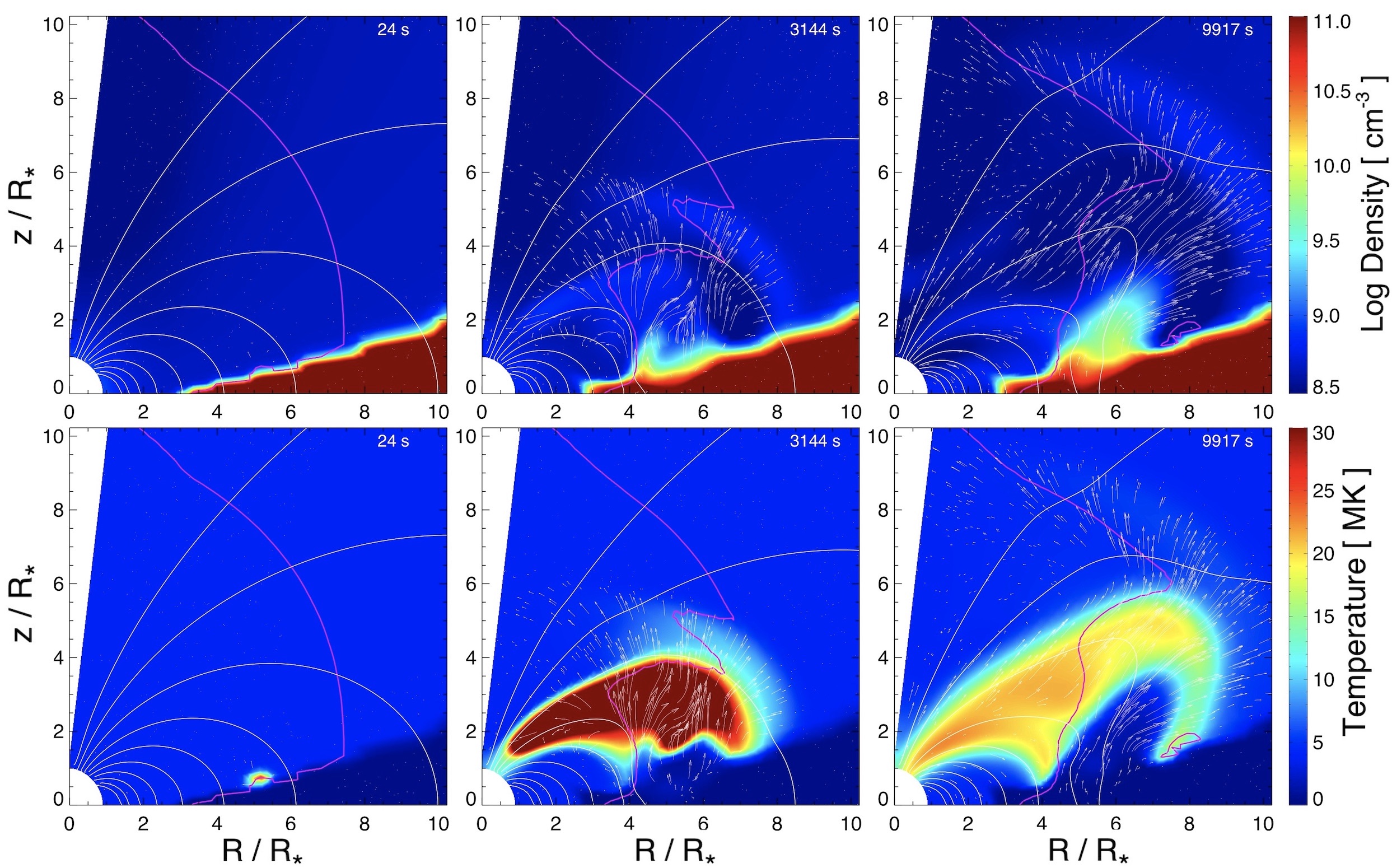}
    \caption{Development of the thermal density (top three panels) and temperature (bottom three) of the system at 24, 3144 and 9917 seconds after the heat pulse is initiated.  The formation of a hot flaring loop is seen to occur between 24 and 3144s.}
    \label{fig:fig4}
\end{figure}

We use several time shots of data, some while the heat pulse is still active but most after it is switched off.
Fig.~\ref{fig:fig4} shows how the flaring loop develops and evolves. The figure shows the distributions of density (upper panels) and temperature
(lower panels) on planes perpendicular to the XY plane and passing
through the middle of the flaring loop at 24, 3144, and 9917 seconds.
The lines represent sampled poloidal magnetic field lines and the
arrows describe the velocity field. The magenta line delimits the
region with plasma $\beta < 1$ on the left of each panel. The heat pulse is confined to a small area (see panels
on the left of Fig.~\ref{fig:fig4}), compared to the loop (middle and right
panels of Fig.~\ref{fig:fig4}). This simulated flaring loop proved to match
analysis from observations well, in terms of its size, maximum
temperature, and peak X-ray luminosity \citep{favata2005bright, orlando2011mass}. 
As the heat is conducted to the disk, part of the local disk material evaporates in the outer stellar atmosphere.  A small fraction of this material is
channelled into the loop and accretes onto the star; however, most of this evaporated material is not confined by the magnetic
field and propagates away from the central star in the outer stellar
corona (see the velocity field in Fig.~\ref{fig:fig4}). Furthermore the flare
strongly perturbs the disk: an overpressure develops in the disk
at the loop's footpoint and travels through the disk, reaching its
opposite side in few hours. When this happens, a funnel flow starts
to develop there, accreting substantial disk material onto the star
from the side of the disk opposite to the flare.

We focus our analysis during the first phase of evolution, namely when the hot magnetic loop develops. Next, we describe how non-thermal electrons (which are not included within the MHD model) are treated.

\subsection{Non-thermal electron model}
\label{subsec:fieldline}

In order to model the non-thermal electron distribution, we assume that particle acceleration (associated with magnetic reconnection) is cospatial and cotemporal with the heating. The accelerated electrons are trapped in a closed magnetic flux tube. This flux tube is comprised of the field lines threading the flaring region. The non-thermal electrons quickly spread along the flux tube and are eventually lost. We do not model the processes by which these particles are lost (e.g. collisional scattering), instead assuming a phenomenological constant decay time.


In order to model the emission from non-thermal electrons, we first identify the energy release region, then trace the magnetic field and determine the flaring loop, i.e. the volume magnetically-connected to the acceleration region (denoted by the location of the heat pulse). Then the flaring loop is populated with energetic electrons. This flux tube is anchored to the stellar surface at two points (its footpoints) and has its apex within the heat pulse region. 


The field lines connected to this region are traced and selected only once, at the start of the simulation. We assume that the magnetic field is "frozen" on the stellar surface and, hence, the footpoints locations remain the same. Therefore, at later times the flux tube volume is identified by tracing the magnetic field from the footpoints. Whilst the star is in fact rotating, the rotation period (9.2\,days) is far greater than the simulation time (approximately 9 hours). 

There is a strong convergence of the field between the loop's apex and stellar surface, leading to a very small loss cone angle. Also, as we are using results from ideal MHD simulations any electric field effects are negligible. Thus, the main effects to consider are magnetic mirroring and collisional losses. In deciding upon the spatial distribution for the non-thermal electrons, a simple Gaussian function was used:
\begin{equation}
\label{eq:gaussian}
n = n_{\rm o} e^{-\frac{l^{2}}{s^{2}}}
\end{equation}

where $n$ is the non-thermal particle density (cm$^{-3}$), $n_{\rm o}$ is the loop top density, $l$ is the distance from the loop top along the flux tube and $s$ is the confinement length scale.  Such a distribution is used for modelling solar flare non-thermal electrons \citep{fleishman2018revealing}. Varying the parameter $s$ controls the distribution of particles throughout the flux tube, as discussed in Section~\ref{subsubsec:nth}. 

\begin{figure}
	\includegraphics[width=\columnwidth]{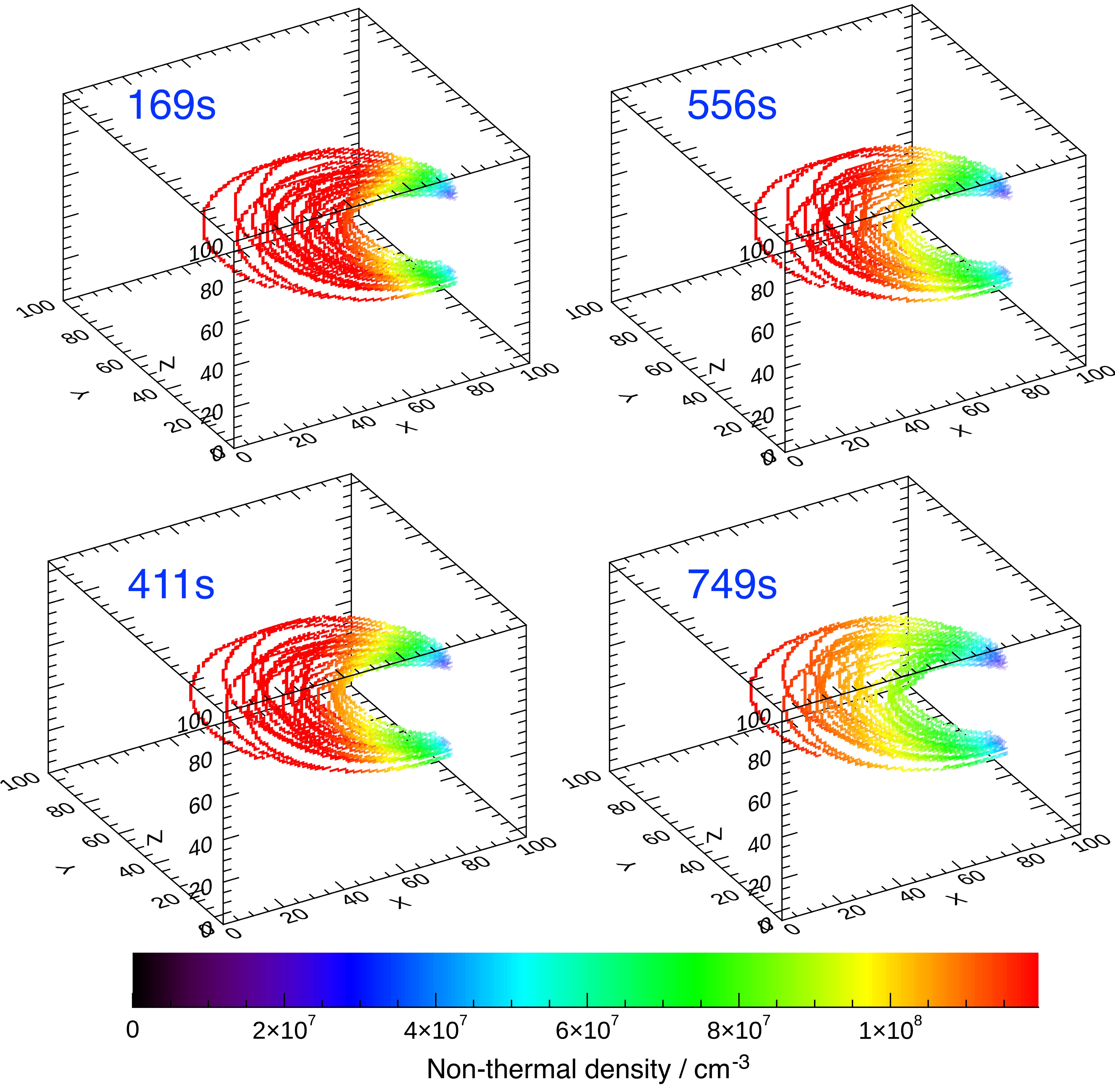}
    \caption{3D visualisation of sample flux tube field lines and how the density of the non-thermal particles vary along them, from 169 to 749s. The non-thermal density throughout the tube is measured in cm$^{-3}$ and is highest at the top of the loop. The flux tube changes shape and position over time, but remains anchored at the same footpoints throughout.}
    \label{fig:96volume}
\end{figure}

As the injection site of the electrons is at the loop top in the high temperature region, this is where the density peaks. A value of 10 percent of the thermal density at that same location was chosen for the initial value of the non-thermal electron density. A 3D visualisation of sample field lines that comprise the flux tube is shown in Fig.~\ref{fig:96volume}.  The non-thermal electrons filling the flux tube have a density distribution shown by the colour scale along these field lines, i.e. the highest particle density is at the loop apex. The flux tube is largely obscured along the X and Y axis by the circumstellar disk but is more visible along the Z line of sight, looking down onto the disk.  The density decreases to a minimum near the footpoints. 

The non-thermal electrons are assumed to have an energy distribution defined by a single power law: 
\begin{equation}
\label{eq:powerlaw}
n(\epsilon)d\epsilon = A\epsilon^{-\delta} d\epsilon
\end{equation}
for ${\epsilon_0 < \epsilon < \epsilon_{max}}$ where $A$ is a normalisation constant.  The energy range used was 0.1MeV< $\epsilon$ < 10MeV. We have chosen the higher than default (as in \citet{Nita2015}) $\epsilon_0$ value to reflect the higher plasma temperatures observed from these sources compared with solar flares. Initially, the power law index, $\delta$, is set at 3.2 throughout the simulation. The electron pitch angle distribution is assumed to be isotropic. 

Despite the effect of magnetic mirroring trapping the particles, the density will eventually decline with time. The injection site of the non-thermal particles is located at the heat pulse site, i.e. the loop apex where magnetic reconnection is assumed to take place. As the particles gyrate along the field lines they move in the direction of the footpoints anchored to the stellar surface.  The approximate relativistic travel time of a 1.0 MeV electron along a half loop length is $\approx$ 18 seconds. It is therefore reasonable to assume that the electrons have time to propagate close to the footpoints.  However, due to the small loss cone angle it is expected that these particles are initially mirrored and remain on the field lines. This is reinforced by examining the approximate trapping time of these electrons. The trapping time \citep{aschwanden1997electron} of a 1.0 MeV electron within a fully ionised plasma (with temperature and density equal to those used in this model) is $\approx$ 2000 seconds. Therefore it is assumed the electrons are magnetically trapped and are able to bounce back and forth multiple times before eventually being lost, either to space or impinging on the stellar surface.  To parameterise this, an exponential time decay component is added after 300s of the form:
\begin{equation}
\label{eq:timeconst}
e^{-\frac{t-300}{t_{\rm o}} }
\end{equation}

where $t$ ($>$ 300s) is the time (in seconds) during the simulation and $t_{\rm o}$ has some constant value representing the trapping or decay time.  Initially $t_{\rm o}$ is set as 2000s to ensure all non-thermal particles have decayed away by the end of the simulation.  This parameter is varied later on in Section~\ref{subsubsec:time} to see its effect on the light curves.  

The loop top non-thermal density decreases for all subsequent times, as there is no further electron acceleration after the explosive energy release due to magnetic reconnection in the first 300s. This is seen in Fig.~\ref{fig:96volume}, with the peak non-thermal density dropping in magnitude as time progresses. 

\subsection{Fast GS simulations}
\begin{figure}
	\includegraphics[width=\columnwidth]{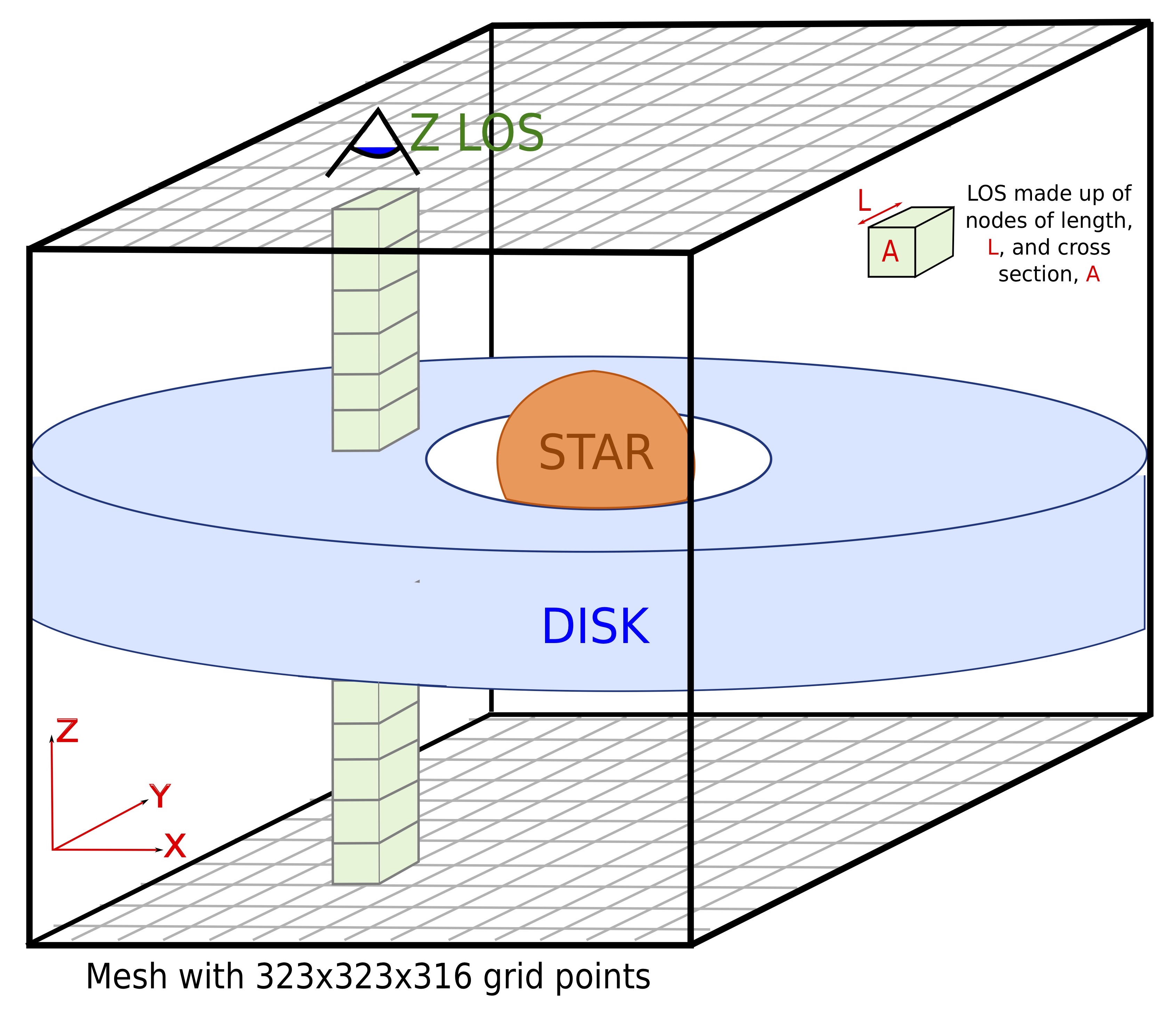}
    \caption{Schematic of how the 1D fast GS simulations work.  The radiative transfer equation is solved along each line of sight, with multiple lines of sight being used to cover the whole field of view. Lines of sight used are in the X, Y and Z direction. The Z line of sight (shown here) looks down onto the disk while the X and Y axis are orientated to look through the disk edge on.}
    \label{fig:LOS1}
\end{figure}

We use fast GS simulations, based on the fast gyrosynchrotron codes developed by \citet{fleishman2010fast}, to calculate the gyrosynchrotron emission from this flaring event over time.  The fast GS simulations calculate the emission and corresponding spectra for 1D lines of sight. Each individual line of sight consists of a predetermined number of nodes (100 in this case) with a specific cross section and length passing through the MHD simulation grid, see Fig.~\ref{fig:LOS1}. We use these GS simulations to calculate the emission in each of the X, Y and Z planes, integrating multiple lines of sight to cover the whole region. 

The data from \citet{orlando2011mass} gives us the input for temperature, thermal density and magnetic field, see Section~\ref{subsec:mhd}.  From these, the non-thermal density is also calculated (discussed in detail in Section~\ref{subsec:fieldline}) as well as the angle between the magnetic field and line of sight. The electron energy distribution is set as a single power law with an isotropic distribution in pitch angle. This is generally a good description of solar flares \citep{Benz2008}. 

\section{Results}
\label{sec:results}

\subsection{Initial results}
\subsubsection{No added non-thermal particles}

First, we explore the case of thermal gyrosynchrotron emission that is calculated only from the MHD model of the flaring star. The thermal density in the flaring region at early times is dwarfed by that of the dense accretion disk.  However at later times the density from the flaring region is comparable to that of the surrounding disk and a clear loop shape forms (see lower panels of Figure~\ref{fig:fig4}).  Overall, the thermal emission spectra for all times is dominated by the emission from the large accretion disk. 

\begin{figure}
	\includegraphics[width=\columnwidth]{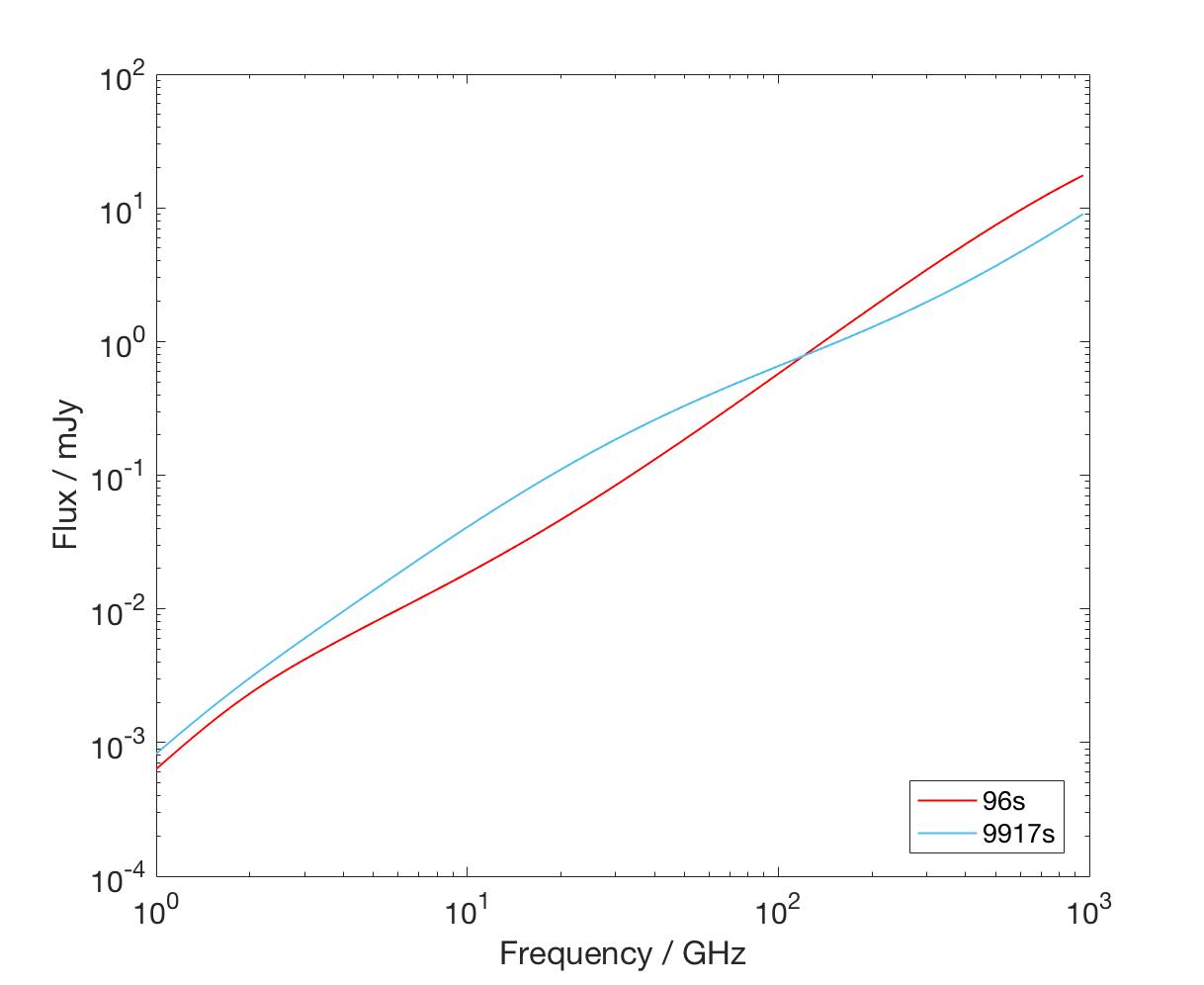}
    \caption{Spectra for the thermal case, at 96 and 9917 seconds after the heat pulse is initiated.  The total intensity (measured in mJy for a distance of 140pc) for all Z lines of sight in the domain is calculated and plotted over a frequency range of 1-1000\,GHz.}
    \label{fig:zthermal96}
\end{figure}

Spectra were generated covering the domain where the flux tube established in Section~\ref{subsec:fieldline} was not populated with any non-thermal particles.  The disk is still present and the evolution of the system as a result of the flaring heat pulse is observed.  There are still thermal particles present (not constrained to field lines) and fluctuating over time in the vicinity of the heat pulse, as well as variations in the temperature. Fig.~\ref{fig:zthermal96} shows the spectra for this thermal system at 96s and 9917s seconds after the heat pulse is switched on.  At 96s, the total intensity increases over 4 orders of magnitude over the 1000\,GHz range of frequencies.  The spectral profile rises gradually with no peak.  

In general, the 9917s spectrum has a similar shape and intensity as for the 96s case. Its flux rises steadily over several orders of magnitude between 1 and 1000GHz. The intensity between the two spectra differ at most by approximately 0.15mJy at 50\,GHz, due to the changing thermal density and distribution over time (Figure~\ref{fig:fig4}). In comparison, Figure~\ref{fig:YLOSSPEC} shows how the added non-thermal electrons lead to a more dramatic change in intensity of approximately 5mJy at 50\,GHz between the same time steps.

The thermal spectra for the remaining time steps are similar to those shown above so are not shown here. The same is true for the respective spectra along the X and Y lines of sight.

\subsubsection{Emission from non-thermal particles}
The next step in exploring the gyrosynchrotron emission from this modelled event is the inclusion of non-thermal particles.  It has been shown that the electrons rapidly fill the flux tube, extending down to the footpoints where they are magnetically trapped.  The gyrosynchrotron emission resulting from populating this flux tube, and the effect of the injected non-thermal particles over time are now explored.  

\begin{figure}
	\includegraphics[width=\columnwidth]{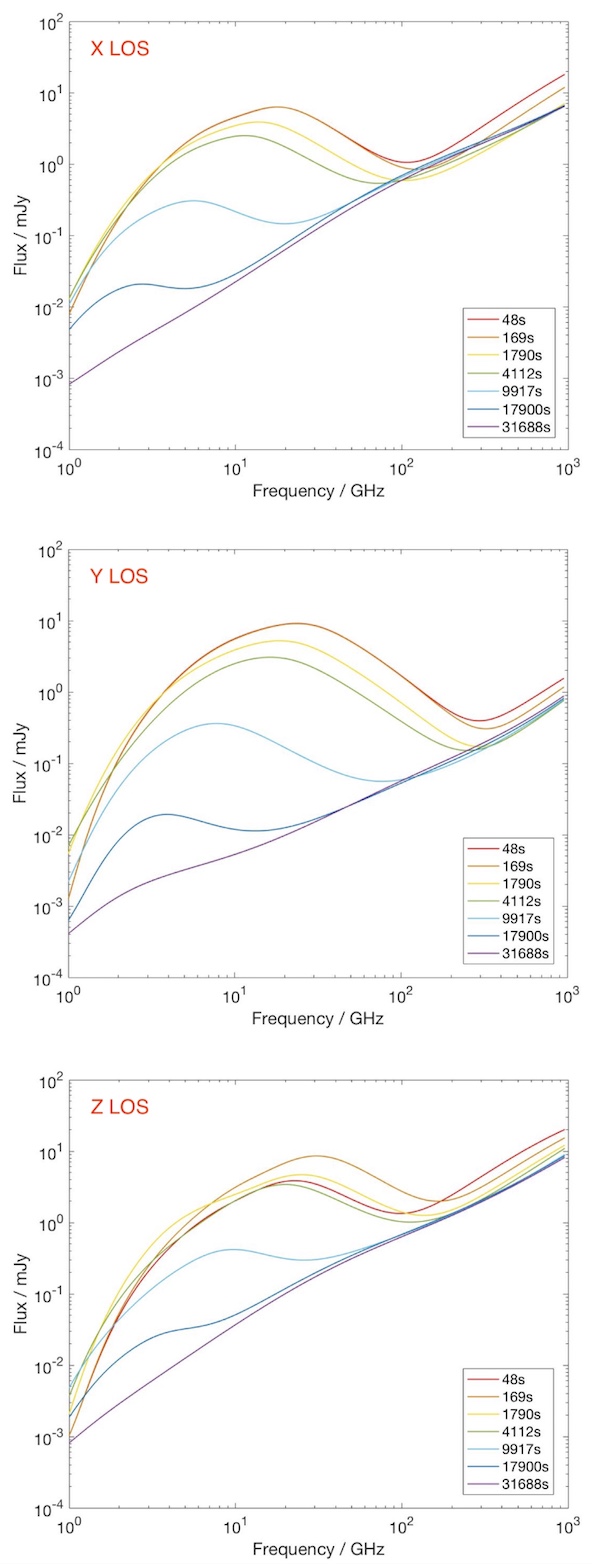}
    \caption{Spectral results for gyrosynchrotron emission generated along the X, Y and Z axes for the whole region. The total intensity is measured in mJy, and is calculated for a frequency range of 1-1000\,GHz at a distance of 140pc.  By the end of the simulation the spectra has relaxed to a nearly thermal state with no influence from the non-thermal particles.  The different coloured lines correspond to successive times, given in seconds from 48 to 31688s.}
    \label{fig:YLOSSPEC}
\end{figure}

Fig.~\ref{fig:YLOSSPEC} shows spectra generated at seven different times and over the three lines of sight. The model results have been scaled to give the flux of the source at a distance of 140pc, approximately that of Taurus \citep{mooley2013b}. Starting at the earliest time, 48 seconds after the heat pulse is started, the flux tube is chosen to be comprised of the field lines passing through the region of highest temperature.  The non-thermal particles are distributed in the loop as described in Section~\ref{subsec:fieldline} with the loop top non-thermal density chosen as ten percent of the thermal density at that same point. Outside of the volume filled by this flux tube, the non-thermal density is zero. This distribution remains constant until 300s onwards, when the number density along the loop is allowed to decay exponentially. This time decay constant is chosen so that by the last time shot the non-thermal density is insignificant enough that the spectra relaxes back to the thermal spectra seen in Fig.~\ref{fig:zthermal96}. The peak non-thermal loop top density, with this baseline parameter set, occurs at 169s with a value of 1.78$\times$10$^8$ cm$^{-3}$.

Looking at the top (X LOS) panel in Fig.~\ref{fig:YLOSSPEC}, the addition of non-thermal particles leads to a `bump' in the spectra around 20\,GHz.  The maximum peak occurs at 169s with a total intensity of 6.33 mJy (at a distance of 140pc) and corresponding frequency of $\approx$18\,GHz. However the spectra for 48 and 169s have very similar profiles and peak values below 100\,GHz. As time progresses the total intensity and frequency of the peak decline.  Near the end of the simulation the non-thermal bump is small and occurs around 2\,GHz with the spectra in general resembling that of a purely thermal case. So, in general, the fewer non-thermal particles present (e.g. at later times after the particles have decayed away), then the lower the peak flux and frequency.  


At frequencies greater than 100\,GHz the spectra all have a similar shape, with less influence from the non-thermal particles.  At 1000\,GHz the highest total intensity is at 48 seconds, decreasing with increasing time. This highest total intensity is 18.1 mJy. At the lowest frequencies (around 1\,GHz) the lowest total intensity is found for the latest, lowest non-thermal density, time shot at 0.823 mJy. 

These patterns are similar for both the X and Y lines of sight in Fig.~\ref{fig:YLOSSPEC}. The main differences between the lines of sight are in the total intensities at relevant frequencies.  For the Y LOS, the highest peak occurs at 48 seconds at 9.12 mJy and $\approx$24GHz.  Unlike for the X LOS, this is the highest total intensity across all frequencies.  The highest total intensity at 1000\,GHz is 1.55 mJy.  The pattern of decreasing peak frequency for increasing time remains the same. The low frequency total intensities are of a lower magnitude than for the X line of sight also (0.408 mJy at 1\,GHz for 31688s).

\begin{figure}
	\includegraphics[width=\columnwidth]{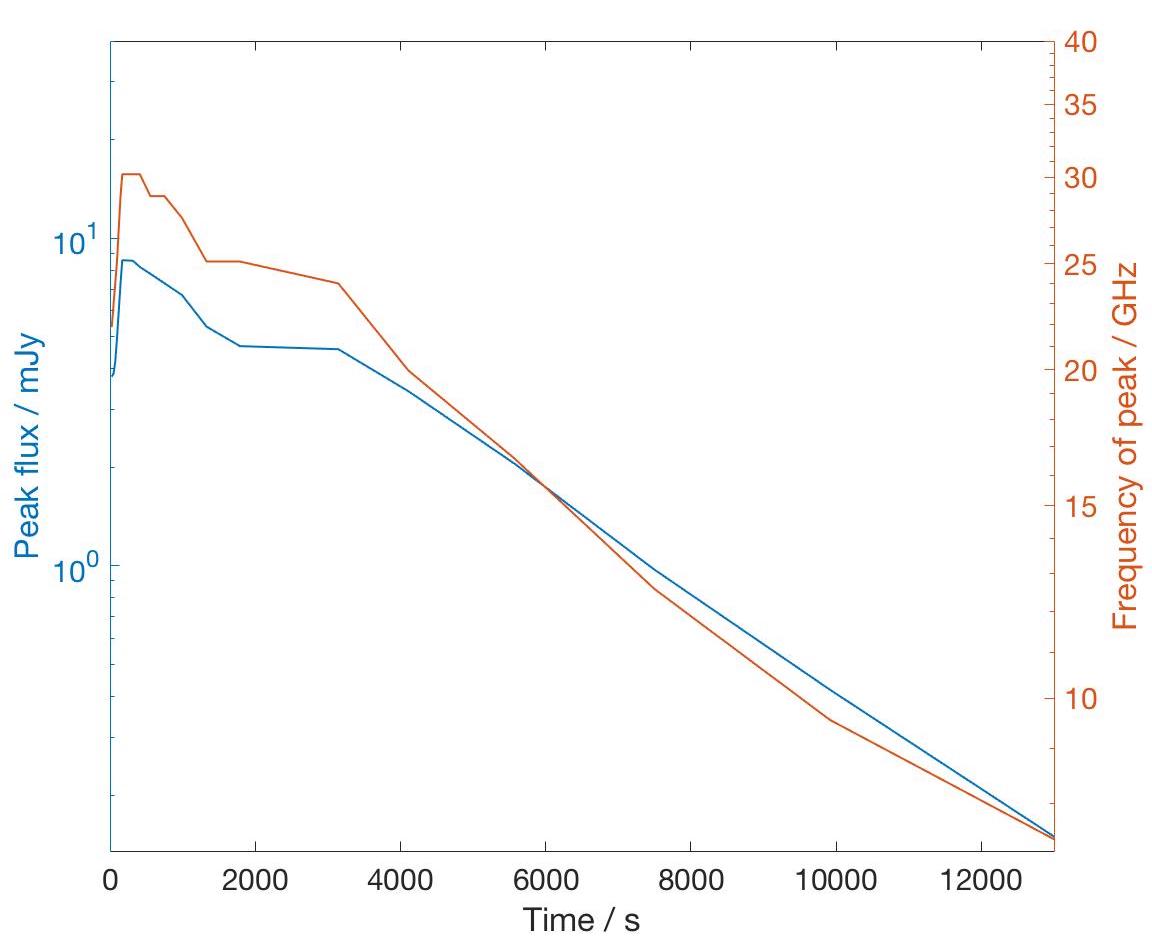}
    \caption{Plot of peak flux of the gyrosynchrotron `bump' in mJy (left axis, blue) and corresponding frequency (right axis, red), over time (in seconds) along the Z LOS. }
    \label{fig:peaks}
\end{figure}

The main factor causing the differences in these spectra is the different magnetic field orientation relative to the line of sight. The final panel in Fig.~\ref{fig:YLOSSPEC} shows the spectra for the Z line of sight at different times. Again, the overall pattern remains similar to that discussed previously.  There is a non-thermal influenced bump around 20\,GHz with the spectra rising to higher total intensities past 100\,GHz.  The highest non-thermal induced peak occurs at 8.57 mJy and $\approx$30\,GHz, at 169 seconds.  As time progresses the peak flux drops and moves to a lower frequency once the heat pulse has been switched off.  This is illustrated in Fig.~\ref{fig:peaks}.  The top plot show this trend over all times for the Z line of sight. There is an increase in flux and peak frequency as the heat pulse evolves up to 300s which then drops and flattens out more at later times. 

\begin{figure}
	\includegraphics[width=\columnwidth]{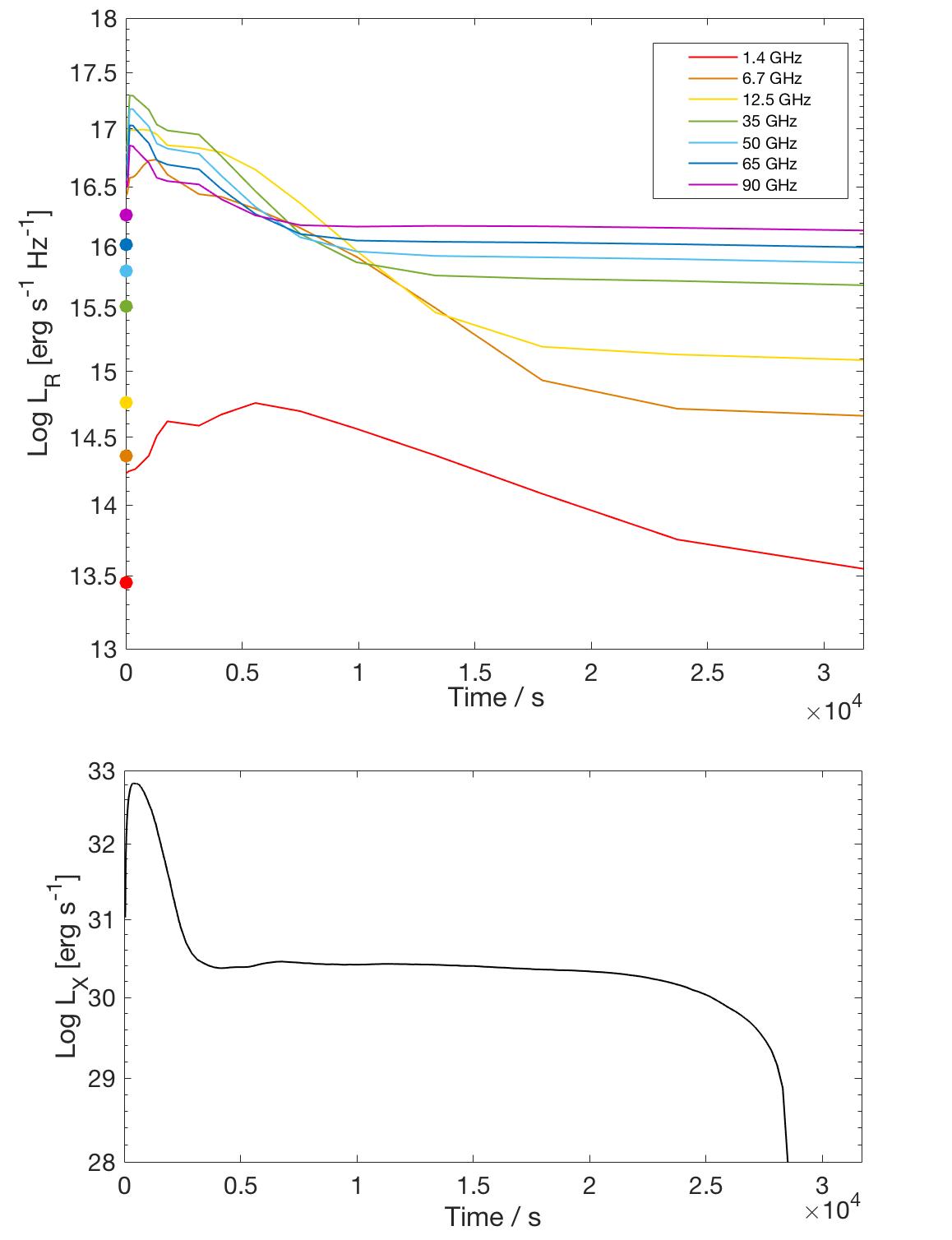}
    \caption{Top: Radio light curves at different frequencies (solid lines). The different colours correspond to different frequencies. The time range is 24-31688 seconds. The pre-flare background luminosities at these frequencies are given as the solid points at $\textrm{t}=0$. LogL$_\textrm{R}$ (erg s$^{-1}$ Hz $^{-1}$) $\approx$ 2.35$\times$10$^{16}$ $\times$ [mJy flux value at 140pc]. Bottom: X-ray light curve from \citet{orlando2011mass}.}
    \label{fig:XLOStime}
\end{figure}

It is also useful to consider the light curves in specific frequencies, showing how intensity varies with time. The results of these for the Z LOS are shown in Fig.~\ref{fig:XLOStime}.  The corresponding X-ray light curve (0.6-12 keV band) from \citet{orlando2011mass} is shown in the second panel. The frequencies used in the radio light curves correspond to the lower frequency observing bands on ALMA \citep{wootten2009atacama, huang2016atacama}  and the bands planned for SKA \citep{braun2017anticipated}. The radio light curves are given in terms of their luminosity, so as to be easily comparable with the G\"udel-Benz relationship discussed in Paper I \citep{waterfall2018modelling}. At early times, the radio luminosity peaks in higher frequency bands. Specifically, the highest radio luminosity occurs for 35\,GHz, which is much higher than these flaring stars are currently typically observed at (most are taken below 10\,GHz, for example: \citep{Osten2009, Dzib2013}) . As the peak luminosity drops, the frequency of this peak also decreases as time progresses.  Similar patterns are seen for the X and Y LOS. 

At late times, when there are few non-thermal particles remaining, the highest luminosities are observed at the highest frequencies because it is dominated by thermal emission. However, the change in luminosity at these higher frequencies (e.g. 90\,GHz) over all time is smaller when compared with the 35\,GHz changes. 

The markers at $\textrm{t}=0$ in Figure~\ref{fig:XLOStime} are the pre-flare background luminosities at each frequency. This background is not subtracted from the spectra so that the spectra more closely represent the flux expected in observations. It should be noted that the simulation period considered here (approximately 9 hours) is less than the original MHD simulation (48 hours). Consequently, the system has not completely relaxed to its pre-flare state at some frequencies, e.g. at 1.4GHz.

The X-ray luminosity (taken from \citet{orlando2011mass}) of this event peaks around 500s, after the heat pulse is switched off.  In contrast, the radio luminosity peaks at 169s, while the flare is still active. This result agrees with patterns observed in solar flares \citep{Benz2008}.  During the impulsive phase, the microwaves peak and drop off sharply. During the secondary, decay phase the thermal X-rays reach their peak. Figure~\ref{fig:XLOStime} shows how the X-ray luminosity decreases more rapidly than the radio luminosity at first.   The flare itself is only partially confined to the magnetic field in the MHD simulation. This causes a large and rapid loss in accreting disk material that is not trapped in the generated star-disk loop. The most relevant characteristic of this X-ray spectrum is the value of the peak itself.  This peak luminosity, as stated by \citet{orlando2011mass}, is comparable to those observed in the brightest of X-ray flares on young stars. 

As discussed in Paper I, the radio luminosity of these flaring stars is higher than expected when compared with the G\"udel-Benz relation. From Fig.~\ref{fig:XLOStime} the radio luminosity peaks at logL$\rm _R$ (erg s$^{-1}$ Hz $^{-1}$) = 17.3, while the corresponding X-ray luminosity peak places the model just below the G\"udel-Benz relation (seen in Paper I). 

The radio luminosities obtained so far are consistent with observational results. However, there are certain parameters within the model which can be varied to see their effect. These include the distribution of particles within the loop, the decay rate of the non-thermal electrons and the power law index.  The results of varying these are discussed next. 

\subsection{Varying the parameters of the non-thermal electrons}
\label{subsec:vary}
The complete equation governing the distribution and loss of electrons in the flux tube is (combining Equation~\ref{eq:gaussian} and Equation~\ref{eq:timeconst}):

\begin{equation}
n = n_{\rm o} e^{-\frac{l^{2}}{s^{2}}} \cdot e^{-\frac{t-300}{t_{\rm o}} }
\end{equation}

In this section $s$, $t_\textrm{o}$ and $\delta$ (power law index; see Equation~\ref{eq:powerlaw}) are all varied from their standard values ($s$=1.0, $t_\textrm{o}$=2000 and $\delta$=3.2)  All the resulting spectra and light curves are given for the Z LOS, unless otherwise specified. 

\label{subsec:Bfield}

\subsubsection{Power law}
\begin{figure}
	\includegraphics[width=\columnwidth]{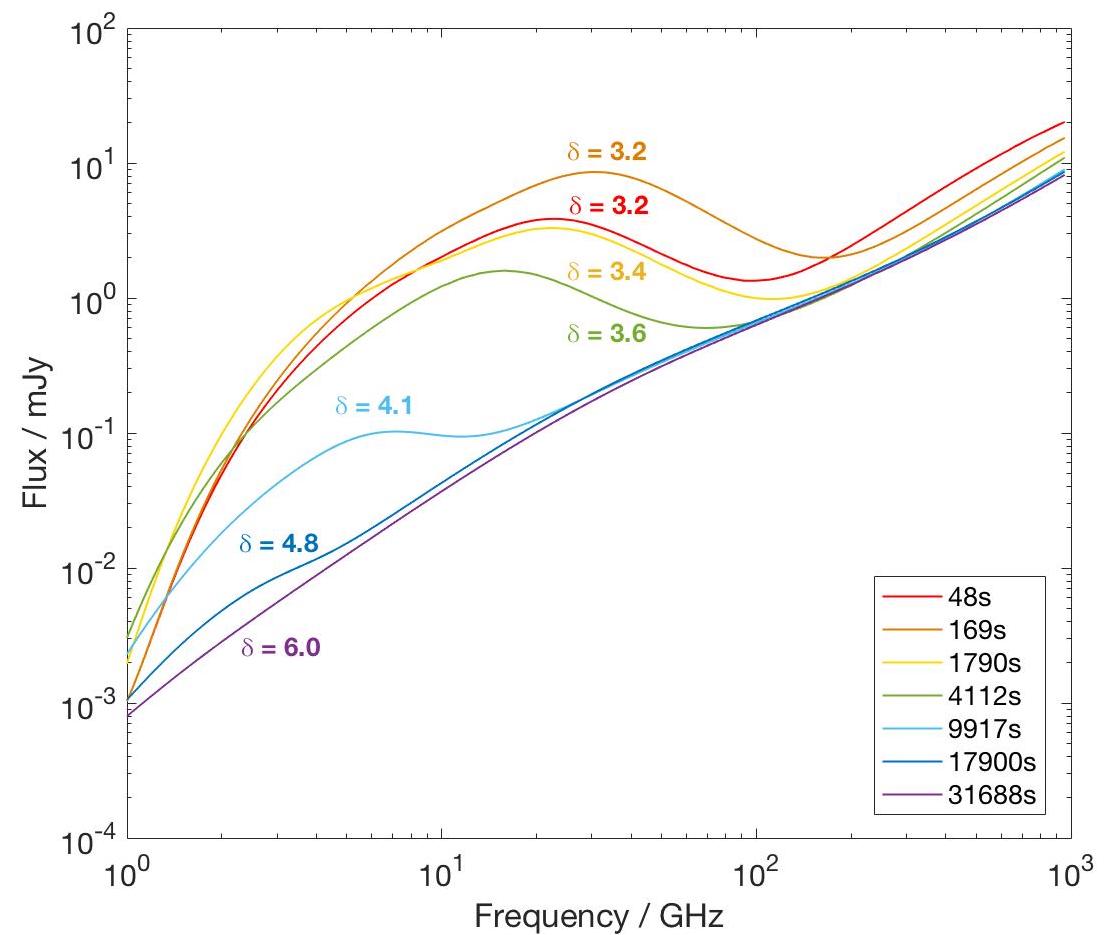}
    \caption{Results of increasing the power law index of non-thermal particle energy over time on the Z LOS spectra.  Time steps are the same as for previous spectra, index values used are labelled for each spectrum.}
    \label{fig:powerlaw}
\end{figure}

The first parameter that was varied was the power law index (Equation~\ref{eq:powerlaw}). Initially, the power law index was set at $\delta$=3.2 for all times (the same value used in Paper I).  As the non-thermal particles are lost, the particle spectrum will become steeper over time \citep{Benz2008}. To model this we allowed $\delta$ to increase linearly over time to reach $\delta$=6 by the final time step. Overall, the number of non-thermal particles decays over time, as before, but the gyrosynchrotron emission is affected by the spectrum of the non-thermal electrons as well as number. Hence, we model the effect of allowing $\delta$ to change. The result of this increasing $\delta$ is a faster decrease in peak flux of the radio spectra over time, see Fig.~\ref{fig:powerlaw}.  This effect can also be seen in the simulation of solar flares by \citet{fleishman2020decay}, movie S2. As they increase the power law index, the spectral peak moves to a lower intensity.

When there are no non-thermal particles present, the thermal spectrum is generated from mostly low energy particles. The inclusion of non-thermal electrons with a single power law of index $\delta=3.2$, gives a spectral peak seen around 20\,GHz,  due to the addition of higher energy electrons to the system.  The smaller power law indices represent a larger proportion of higher energy particles which represents the larger number of non-thermal electrons present at the start of the simulation when compared with the end.  As time progresses, this power law index increases and the energy distribution becomes steeper, limiting the number of higher energy electrons. This leads to spectra becoming more `thermal' in shape, with the peak generated from the non-thermal particles flattening out. This is seen in the spectra of Fig.~\ref{fig:powerlaw} compared to the Z LOS spectra from Fig.~\ref{fig:YLOSSPEC}. At earlier times, the peaks are similar, with the power law index not increasing noticeably until later times where the system relaxes to a thermal state much faster than when the index was a constant value. The addition of an increasing power law index therefore results in the spectral peak being reduced faster and pushed to lower frequencies as the contribution of higher energy particles is reduced.  

\subsubsection{Distribution along the loop}
\label{subsubsec:nth}

The distribution of non-thermal particles within the loop is defined by Equation~\ref{eq:gaussian}. As described in Section~\ref{subsec:fieldline}, $l$ is the distance from the loop top to one footpoint, or the loop half length. The factor which controls the distribution within the loop, $s$, is initially set as this half loop length.  As is seen in Fig.~\ref{fig:96volume} the density is always highest at the apex of the loop, dropping off to a minimum at the footpoints. 

\begin{figure}
	\includegraphics[width=\columnwidth]{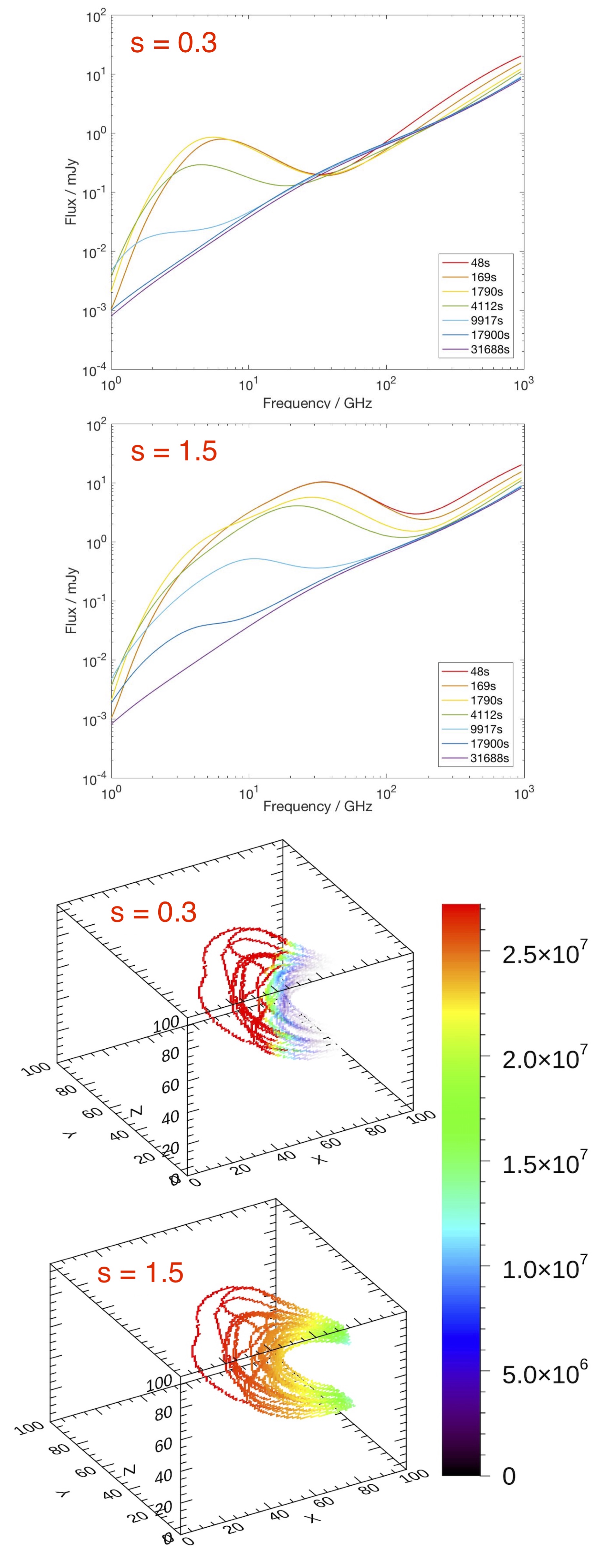}
    \caption{Spectra along the Z LOS resulting from changing the non-thermal electron density spatial distribution (top two plots). Two different configurations ($s$ = 0.3 and 1.5) are considered over the seven time steps. The results for $s$ = 1.0 are seen in the Z LOS spectra of Fig.~\ref{fig:YLOSSPEC}. Bottom images are 3D visualisation of sample flux tube field lines at 4112s and their electron distributions. Colour bar units are cm$^{-3}$.}
    \label{fig:dist}
\end{figure}

The parameter $s$ was varied to investigate the effect on spectra of different spreads in non-thermal electrons within the loop. The results from using two values of $s$ are shown in Fig.~\ref{fig:dist}. The top two panels show the spectra along the Z LOS for $s$ parameter values of 0.3 times the half loop length (same as the Z LOS spectra in Fig.~\ref{fig:YLOSSPEC}) and 1.5 times the half loop length. 

Also shown in Figure~\ref{fig:dist} is a 3D visualisation of these two different distributions along the field lines (and throughout the total flux tube) at 4112 seconds (bottom two panels of Fig.~\ref{fig:dist}). For a smaller value of $s$, the particles are more constrained to the loop top while a larger $s$ value leads to a more even distribution from the apex down towards the foot points.  The point of highest non-thermal density is found in the ${s = 0.3}$ plot as the total mass is conserved before the time decay component is added (Equation~\ref{eq:timeconst}). 

For the more compact distribution (${s = 0.3}$) in Fig.~\ref{fig:dist} it is clear the system relaxes to a thermal state quicker than for the more extended distribution of electrons and has a lower peak flux overall.  However, the peak in emission occurs at 1790s, later than the 169s seen for the other distributions and lines of sight. Overall, however, the pattern of the spectral evolution is similar. After the peak flux is reached, the decay of the non-thermal particles leads to the decrease in peak flux and frequency at subsequent times. The spectra for $s$ = 1.0 and 1.5 are generally similar in shape. However, the larger concentration of higher density particles along the last loop leads to a higher peak flux at earlier times before the decay of non-thermal electrons starts. These spectra agree with those for solar flares from \citet{kuznetsov2011three}, where a more confined electron distribution towards the loop top results in a lower intensity and lower frequency of the spectral peak. 

\subsubsection{Time decay}
\label{subsubsec:time}
\begin{figure}
	\includegraphics[width=\columnwidth]{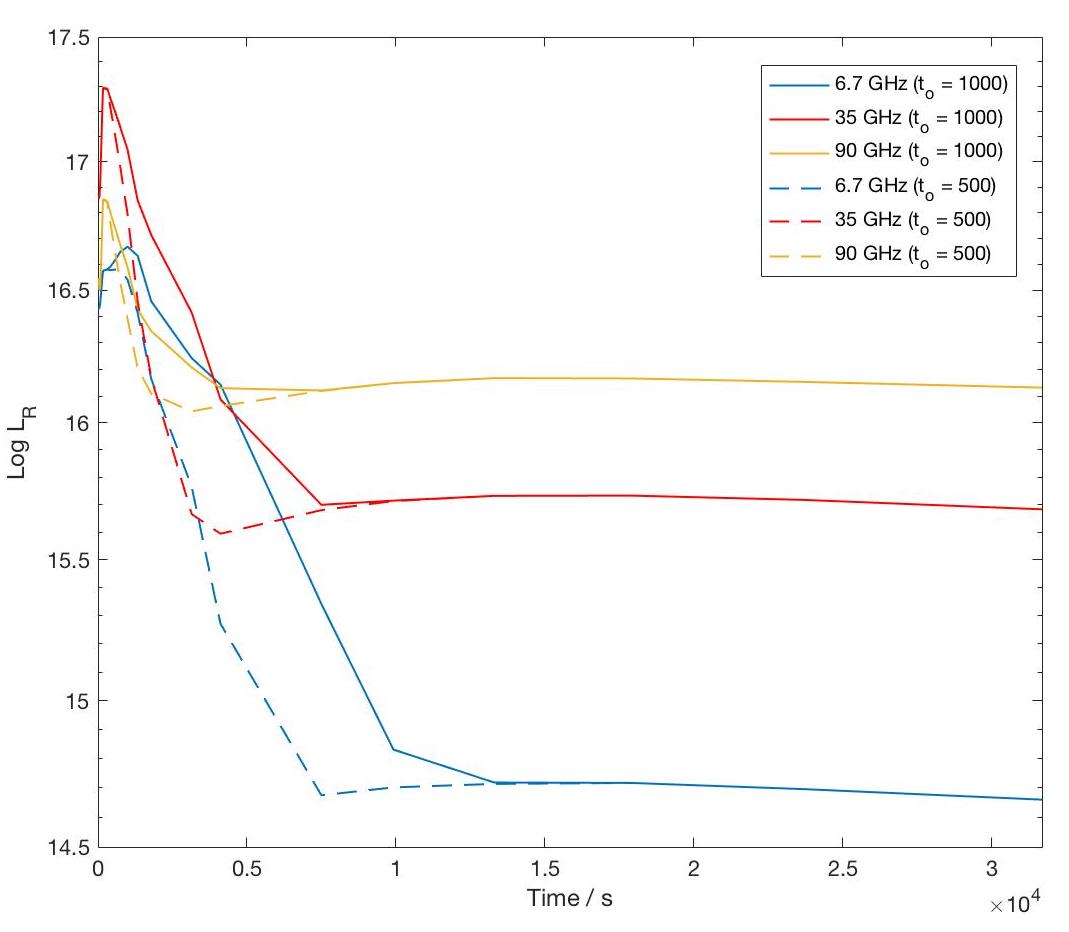}
    \caption{Light curve results for three different frequencies when the time decay constant ${t_{\rm o}}$ is varied. The standard value for previous results is 2000s (see light curve from Fig.~\ref{fig:XLOStime}).  The different ${t_{\rm o}}$ values considered here are 500s and 1000s. A smaller ${t_{\rm o}}$ value leads to the faster loss of non-thermal particles from the system. Radio luminosity units are [erg s$^{-1}$ Hz $^{-1}$].}
    \label{fig:time}
\end{figure}

The exponential time decay of the non-thermal electrons after 300s is given by Equation~\ref{eq:timeconst}.  As was previously discussed, t$_{\rm o}$ is initially set at 2000s. This parameter was varied from 100s to 20000s to consider the effect of varying the loss rate of the electrons. The results of varying this for two different time decay constants (500s and 1000s) are shown in Fig.~\ref{fig:time}.  The three different frequencies for these light curves are taken from Fig.~\ref{fig:XLOStime}, where the light curves for a time constant of 2000s is shown. 

The smaller t$_{\rm o}$ value of 500s leads to a more rapid drop in radio luminosity, indicating the increased loss rate of the non-thermal electrons. The choice of time decay constant has no effect on the frequency at which the emission peaks, with spectra for all values of t$_{\rm o}$ peaking at 35\,GHz as in Fig.~\ref{fig:XLOStime}.   

\subsection{Polarisation}
\begin{figure}
	\includegraphics[width=\columnwidth]{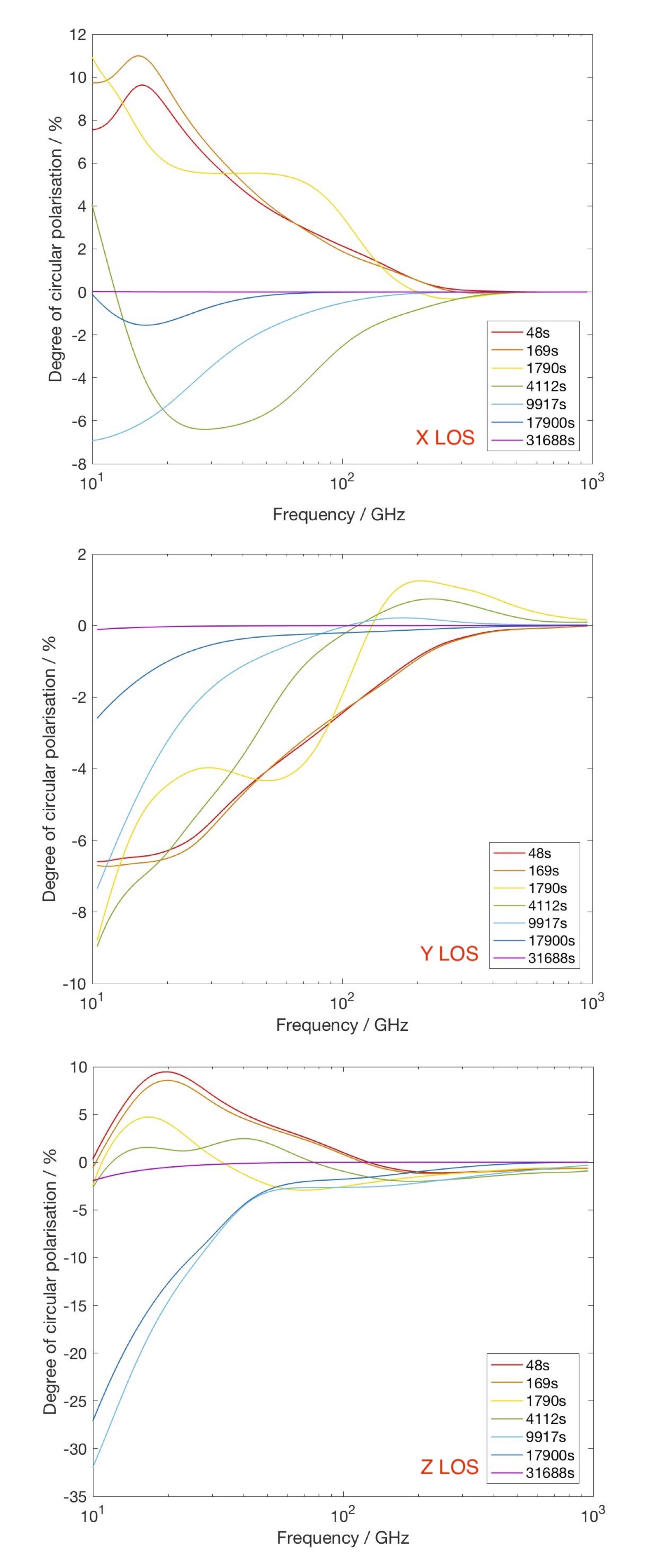}
    \caption{Degree (\%) of circular polarisation at different times over the 10-1000\,GHz frequency range. All 3 LOS are shown.}
    \label{fig:polar}
\end{figure}

One important measurable quantity from observations of stellar flares is the degree of circular polarisation (CP).  Circular polarisation is produced from regions that produce gyrosynchrotron emission from the gyration of mildly relativistic electrons along magnetic field lines.  The circular polarisation detected from a source can vary depending on the field strength and viewing angle, as well as the pitch-angle anisotropy of non-thermal electrons \citep{fleishman2003gyrosynchrotron}. A stronger line of sight magnetic field strength produces a larger degree of circular polarisation. This quantity is therefore closely related to the direction of the magnetic field relative to the line of sight, and potentially provides a diagnostic of the magnetic geometry. The degree of CP is calculated for our fiducial model along all 3 lines of sight over the frequency range 10-1000\,GHz.  The results are shown in Fig.~\ref{fig:polar}.  


A common feature in Fig.~\ref{fig:polar} is the complexity in both time and frequency, which are both also dependent on the viewing angle chosen.  There is a significant degree of CP with a decrease in absolute CP towards zero as time progresses across all three lines of sight. This is due to the reduction in number of non-thermal electrons in time. By 31688s there is zero polarisation over all three LOS as the emission is no longer gyrosynchrotron in nature.  The CP also approaches zero for all three LOS at high frequencies.  This is to be expected from looking at the spectra in Fig.~\ref{fig:YLOSSPEC}.  At high frequencies ($>$ 100\,GHz) there is no contribution to the emission from the non-thermal electrons. 

All three LOS plots display the results in the 10-1000\,GHz range.  The lower frequency emission (below 10GHz) is unpolarised for the X and Y LOS. The Z line of sight however does show some degree of circular polarisation.  This is due to optical depth effects at low frequencies.  In optically thick regions, the sense of polarisation corresponds to either the O-mode (ordinary) or to no polarisation at all. As the X and Y LOS are looking through the dense disk this polarisation becomes zero at these low frequencies. The Z LOS however shows some polarisation in the O-mode as it is looking down upon the disk and unobscured flux tube. 

In optically thin regions however, the sense of polarisation corresponds to the X-mode (extraordinary) so circular polarisation is detectable at these higher frequencies ($>$ 10\,GHz). We have already seen this optically thin region appearing in spectra due to the presence of non-thermal particles. The behaviour of this CP in optically thin regions varies between all 3 LOS however.  Looking at the 3D flux tube visualisations in Fig.~\ref{fig:96volume}, the axis that looks along the magnetic field the most is the Y LOS.  The X LOS looks more in the direction across the field lines while the Z LOS looks down onto them. As the Y LOS is orientated in this way the degree of CP generally does not reverse sign over time.

Both the X and Z LOS have positive CP initially, changing to negative at later times. The change of sign in polarisation can be due to absorption effects or the radiation passing through a region where the field changes direction. However, this effect is also seen looking across flux tubes over time in solar flares. \citet{huang2018solar} reported a change in sign of circular polarisation from the rise to decay phase of a flare.  



\section{Summary and conclusions}
\label{sec:conc}
The variable radio emission from a flaring T-Tauri star has been modelled, based on previous MHD simulations \citep{orlando2011mass}. For the first time, predictions of the multi-frequency intensity and circular polarisation for a T-Tauri star undergoing a flare have been made. The gyrosynchrotron emission and circular polarisation across the X, Y and Z directions of the system have been calculated (for a total simulation time of $\approx$ 9 hours). 

A fast gyrosynchrotron code accounting for emission and absorption due to the thermal and non-thermal plasma is used to calculate the gyrosynchrotron emission from the system for a range of observable scenarios \citep{fleishman2010fast}.  The non-thermal emission originates from a predefined flux tube comprised of a field lines connecting the star to the flaring region near the inner radius of the circumstellar disk. This flux tube is populated with a Gaussian distribution of non-thermal electrons. The flux tube's apex is located at the site of the flare, close to the inner edge of the circumstellar disk. After the flaring heat pulse is switched off, an exponential time loss component is initiated, allowing for the decay of non-thermal electrons over time. 

The fiducial model results are consistent with those from Paper I, as well as with observations.  There still remains the departure from the G\"udel-Benz relation at radio frequencies, even given the lower non-thermal density used compared to in Paper I. The peak radio luminosity from this model is logL$\rm _R$ (erg s$^{-1}$ Hz $^{-1}$) = 17.3 .  In Paper I the peak luminosity from the fiducial model was logL$\rm _R$ (erg~s$^{-1}$ Hz $^{-1}$) =16.3.  These new results further support the idea that these stars produce relatively large amounts of radio emission, associated with their strong magnetic fields, when compared with main sequence stars. 

In Paper I, the physical properties of the system (such as flux tube densities and surface field strengths) were varied.  In this work there are only three varied parameters: the time decay constant (controls the rate at which electrons are lost from the system), the power law index (a similar effect) and the spatial distribution of electrons within the loop. The time decay constant is the least well constrained of these variables. More extensive observations of flares and their variable radio emission will allow tighter constraints on this parameter, and indeed the other parameters in the model.

This paper extends the work done in Paper I by using both an improved model for the magnetic geometry and flaring flux tube as well as including time dependency. As expected, the radio luminosity decreases with time for all frequencies after the heat pulse is switched off.  However, the peak of this luminosity also moves to a lower frequency with time. As more non-thermal electrons are lost, the lower the frequency of the peak in emission becomes. The majority of observations of flaring T-Tauri stars are taken below 10\,GHz, however our model shows the peak fluxes occurring around 30\,GHz. 

The circular polarisation for all three lines of sight across the system and over all times was also calculated.  There is complex behaviour over time and frequency. The degree of polarisation varies considerably, depending on the orientation considered, the frequency, and the time step. This is potentially an important observational diagnostic, but multi-frequency observations of the polarisation, and careful interpretation in comparison with a model are essential. 

The properties of the radio emission predicted by our models are testable through future observations. Observations at higher frequencies are possible with current observatories, however simultaneous, multi-frequency, polarisation observations have not yet been performed. Future observatories such as the ngVLA \citep{mckinnon2018next}, SKA (for lower frequency observations, \citep{braun2017anticipated} as well as Athena for X-ray flares \citep{Barcons2017} will provide opportunities for such observations. A possible next step in modelling these flares includes the use of a resistive MHD model.  Such a flare model, driven by magnetic reconnection, will allow for the simulation of the energy release as well as the trapping and loss of energetic particles.  

\section*{Acknowledgements}
CW is grateful to STFC for studentship support. PB and MG acknowledge STFC support through the consolidated grant ST/P000428/1.



\bibliographystyle{mnras}
\bibliography{paper2}

\begin{thebibliography}{}
\makeatletter
\relax
\def\mn@urlcharsother{\let\do\@makeother \do\$\do\&\do\#\do\^\do\_\do\%\do\~}
\def\mn@doi{\begingroup\mn@urlcharsother \@ifnextchar [ {\mn@doi@}
  {\mn@doi@[]}}
\def\mn@doi@[#1]#2{\def\@tempa{#1}\ifx\@tempa\@empty \href
  {http://dx.doi.org/#2} {doi:#2}\else \href {http://dx.doi.org/#2} {#1}\fi
  \endgroup}
\def\mn@eprint#1#2{\mn@eprint@#1:#2::\@nil}
\def\mn@eprint@arXiv#1{\href {http://arxiv.org/abs/#1} {{\tt arXiv:#1}}}
\def\mn@eprint@dblp#1{\href {http://dblp.uni-trier.de/rec/bibtex/#1.xml}
  {dblp:#1}}
\def\mn@eprint@#1:#2:#3:#4\@nil{\def\@tempa {#1}\def\@tempb {#2}\def\@tempc
  {#3}\ifx \@tempc \@empty \let \@tempc \@tempb \let \@tempb \@tempa \fi \ifx
  \@tempb \@empty \def\@tempb {arXiv}\fi \@ifundefined
  {mn@eprint@\@tempb}{\@tempb:\@tempc}{\expandafter \expandafter \csname
  mn@eprint@\@tempb\endcsname \expandafter{\@tempc}}}

\bibitem[\protect\citeauthoryear{Aschwanden, Bynum, Kosugi, Hudson  \&
  Schwartz}{Aschwanden et~al.}{1997}]{aschwanden1997electron}
Aschwanden M.~J.,  Bynum R.~M.,  Kosugi T.,  Hudson H.~S.,   Schwartz R.~A.,
  1997, The Astrophysical Journal, 487, 936

\bibitem[\protect\citeauthoryear{Barcons et~al.,}{Barcons
  et~al.}{2017}]{Barcons2017}
Barcons X.,  et~al., 2017, \mn@doi [Astronomische Nachrichten]
  {10.1002/asna.201713323}, 338, 153

\bibitem[\protect\citeauthoryear{Benz}{Benz}{2008}]{Benz2008}
Benz A.~O.,  2008, \mn@doi [Living Reviews in Solar Physics]
  {10.12942/lrsp-2008-1}, 5

\bibitem[\protect\citeauthoryear{Braun, Bonaldi, Bourke, Keane  \& Wagg}{Braun
  et~al.}{2017}]{braun2017anticipated}
Braun R.,  Bonaldi A.,  Bourke T.,  Keane E.,   Wagg J.,  2017,
  SKA-TEL-SKO-0000818

\bibitem[\protect\citeauthoryear{Dzib et~al.,}{Dzib et~al.}{2013}]{Dzib2013}
Dzib S.~A.,  et~al., 2013, \mn@doi [The Astrophysical Journal]
  {10.1088/0004-637x/775/1/63}, 775, 63

\bibitem[\protect\citeauthoryear{Favata, Flaccomio, Reale, Micela, Sciortino,
  Shang, Stassun  \& Feigelson}{Favata et~al.}{2005}]{favata2005bright}
Favata F.,  Flaccomio E.,  Reale F.,  Micela G.,  Sciortino S.,  Shang H.,
  Stassun K.,   Feigelson E.,  2005, The Astrophysical Journal Supplement
  Series, 160, 469

\bibitem[\protect\citeauthoryear{Feigelson \& Montmerle}{Feigelson \&
  Montmerle}{1999}]{feigelson1999high}
Feigelson E.~D.,  Montmerle T.,  1999, Annual Review of Astronomy and
  Astrophysics, 37, 363

\bibitem[\protect\citeauthoryear{Fleishman \& Kuznetsov}{Fleishman \&
  Kuznetsov}{2010}]{fleishman2010fast}
Fleishman G.~D.,  Kuznetsov A.~A.,  2010, The Astrophysical Journal, 721, 1127

\bibitem[\protect\citeauthoryear{Fleishman \& Melnikov}{Fleishman \&
  Melnikov}{2003}]{fleishman2003gyrosynchrotron}
Fleishman G.,  Melnikov V.,  2003, The Astrophysical Journal, 587, 823

\bibitem[\protect\citeauthoryear{Fleishman, Nita, Kuroda, Jia, Tong, Wen  \&
  Zhizhuo}{Fleishman et~al.}{2018}]{fleishman2018revealing}
Fleishman G.~D.,  Nita G.~M.,  Kuroda N.,  Jia S.,  Tong K.,  Wen R.~R.,
  Zhizhuo Z.,  2018, The Astrophysical Journal, 859, 17

\bibitem[\protect\citeauthoryear{Fleishman, Gary, Chen, Kuroda, Yu  \&
  Nita}{Fleishman et~al.}{2020}]{fleishman2020decay}
Fleishman G.~D.,  Gary D.~E.,  Chen B.,  Kuroda N.,  Yu S.,   Nita G.~M.,
  2020, Science, 367, 278

\bibitem[\protect\citeauthoryear{Fletcher et~al.,}{Fletcher
  et~al.}{2011}]{Fletcher2011}
Fletcher L.,  et~al., 2011, \mn@doi [Space Science Reviews]
  {10.1007/s11214-010-9701-8}, 159, 19

\bibitem[\protect\citeauthoryear{G{\"u}del \& Benz}{G{\"u}del \&
  Benz}{1993}]{Guedel1993}
G{\"u}del M.,  Benz A.~O.,  1993, \mn@doi [The Astrophysical Journal]
  {10.1086/186766}, 405, L63

\bibitem[\protect\citeauthoryear{Huang et~al.,}{Huang
  et~al.}{2016}]{huang2016atacama}
Huang Y. D.~T.,  et~al., 2016, in Modeling, Systems Engineering, and Project
  Management for Astronomy VI. p. 99111V

\bibitem[\protect\citeauthoryear{Huang, Melnikov, Ji  \& Ning}{Huang
  et~al.}{2018}]{huang2018solar}
Huang G.,  Melnikov V.~F.,  Ji H.,   Ning Z.,  2018, Solar flare loops:
  observations and interpretations.
Springer

\bibitem[\protect\citeauthoryear{Isobe, Shibata, Yokoyama  \& Imanishi}{Isobe
  et~al.}{2003}]{isobe2003hydrodynamic}
Isobe H.,  Shibata K.,  Yokoyama T.,   Imanishi K.,  2003, Publications of the
  Astronomical Society of Japan, 55, 967

\bibitem[\protect\citeauthoryear{Kuznetsov, Nita  \& Fleishman}{Kuznetsov
  et~al.}{2011}]{kuznetsov2011three}
Kuznetsov A.~A.,  Nita G.~M.,   Fleishman G.~D.,  2011, The Astrophysical
  Journal, 742, 87

\bibitem[\protect\citeauthoryear{Long, Romanova  \& Lamb}{Long
  et~al.}{2012}]{long2012accretion}
Long M.,  Romanova M.~M.,   Lamb F.~K.,  2012, New Astronomy, 17, 232

\bibitem[\protect\citeauthoryear{McKinnon \& Selina}{McKinnon \&
  Selina}{2018}]{mckinnon2018next}
McKinnon M.,  Selina R.,  2018, in American Astronomical Society Meeting
  Abstracts \# 231.

\bibitem[\protect\citeauthoryear{Mooley, Hillenbrand, Rebull, Padgett  \&
  Knapp}{Mooley et~al.}{2013}]{mooley2013b}
Mooley K.,  Hillenbrand L.,  Rebull L.,  Padgett D.,   Knapp G.,  2013, The
  Astrophysical Journal, 771, 110

\bibitem[\protect\citeauthoryear{Nita, Fleishman, Kuznetsov, Kontar  \&
  Gary}{Nita et~al.}{2015}]{Nita2015}
Nita G.~M.,  Fleishman G.~D.,  Kuznetsov A.~A.,  Kontar E.~P.,   Gary D.~E.,
  2015, \mn@doi [The Astrophysical Journal] {10.1088/0004-637x/799/2/236}, 799,
  236

\bibitem[\protect\citeauthoryear{Orlando, Reale, Peres  \& Mignone}{Orlando
  et~al.}{2011}]{orlando2011mass}
Orlando S.,  Reale F.,  Peres G.,   Mignone A.,  2011, Monthly Notices of the
  Royal Astronomical Society, 415, 3380

\bibitem[\protect\citeauthoryear{Orlando et~al.,}{Orlando
  et~al.}{2013}]{orlando2013radiative}
Orlando S.,  et~al., 2013, Astronomy \& Astrophysics, 559, A127

\bibitem[\protect\citeauthoryear{Osten \& Wolk}{Osten \&
  Wolk}{2009}]{Osten2009}
Osten R.~A.,  Wolk S.~J.,  2009, \mn@doi [The Astrophysical Journal]
  {10.1088/0004-637x/691/2/1128}, 691, 1128

\bibitem[\protect\citeauthoryear{Robinson, Owen, Espaillat  \& Adams}{Robinson
  et~al.}{2017}]{robinson2017time}
Robinson C.,  Owen J.,  Espaillat C.,   Adams F.,  2017, The Astrophysical
  Journal, 838, 100

\bibitem[\protect\citeauthoryear{Sacco, Argiroffi, Orlando, Maggio, Peres  \&
  Reale}{Sacco et~al.}{2008}]{sacco2008x}
Sacco G.,  Argiroffi C.,  Orlando S.,  Maggio A.,  Peres G.,   Reale F.,  2008,
  Astronomy \& Astrophysics, 491, L17

\bibitem[\protect\citeauthoryear{Sacco, Orlando, Argiroffi, Maggio, Peres,
  Reale  \& Curran}{Sacco et~al.}{2010}]{sacco2010observability}
Sacco G.,  Orlando S.,  Argiroffi C.,  Maggio A.,  Peres G.,  Reale F.,
  Curran R.,  2010, Astronomy \& Astrophysics, 522, A55

\bibitem[\protect\citeauthoryear{Shibata \& Magara}{Shibata \&
  Magara}{2011}]{shibata2011solar}
Shibata K.,  Magara T.,  2011, Living Reviews in Solar Physics, 8, 6

\bibitem[\protect\citeauthoryear{Shibata \& Yokoyama}{Shibata \&
  Yokoyama}{1999}]{shibata1999origin}
Shibata K.,  Yokoyama T.,  1999, The Astrophysical Journal Letters, 526, L49

\bibitem[\protect\citeauthoryear{Waterfall, Browning, Fuller  \&
  Gordovskyy}{Waterfall et~al.}{2018}]{waterfall2018modelling}
Waterfall C.,  Browning P.,  Fuller G.,   Gordovskyy M.,  2018, Monthly Notices
  of the Royal Astronomical Society, 483, 917

\bibitem[\protect\citeauthoryear{Wootten \& Thompson}{Wootten \&
  Thompson}{2009}]{wootten2009atacama}
Wootten A.,  Thompson A.~R.,  2009, Proceedings of the IEEE, 97, 1463

\makeatother
\end{thebibliography}

\bsp	
\label{lastpage}
\end{document}